# PHYSICS OF THE PSEUDOGAP II: DYNAMICS, INCOMPRESSIBILITY, AND FERMI ARCS AND "MOTIONAL NARROWING"

ABSTRACT
A further discussion of the vortex fluid in the cuprate high Tc superconductors is presented. The crucial property of incompressibility towards the addition of net vorticity, leading to the marked nonlinearity of the response functions, is justified from first principles. We also discuss the "Fermi Arc" phenomenon of Campuzano as a consequence of the time-fluctuating phase in the vortex fluid.

It has been known since the work of Salamon in 1993[1], and has been repeatedly confirmed, that cuprate high Tc superconductors show 3Dx-y critical behavior. The necessary implication is that the state immediately above Tc is not a simple metal, dominated by fermionic degrees of freedom, but has a bosonic order parameter with fluctuating phase. The investigations of Orenstein[2] by microwave spectroscopy, and the even more definitive work of Ong and collaborators on Nernst effect and diamagnetism[3], have now confirmed that the region of x-y model behavior extends far above Tc, especially in underdoped materials but to an extent near optimal doping. There is a fairly definite "onset" temperature To, lower than the "pseudogap" temperature T*. Above To there is no evidence for supercurrents, but below it the phenomenology implies fluctuating superconductivity. This behavior also is found to extend to high magnetic fields. It disappears above a certain magnetic field $H_{c2}$ when this can be reached. Whether the onset is a phase transition in the H,T plane remains moot but to me seems likely. (Fig 1)

We have called the rather large volume in the x, T, H phase diagram, where the fluctuating x-y behavior is seen, the "vortex liquid" phase[4]. This terminology is well accepted for the state above the "melting temperature" of the Abrikosov vortex lattice in a magnetic field, and connotes that the vortices caused by the magnetic field have become free to fluctuate; but if we are to apply it also in zero field we must assume that there are thermally activated vortex loops as well. The experiments of Ong, illustrated in his contour plot ( Fig 2) unequivocally show that the vortex liquid is a continuum in the H,T plane.

An argument due to Ong and Wang [5] seems to ensure that the phenomena observed are due to vortices caused by the magnetic field. They show that if and only if this is the case, the Nernst coefficient $\alpha_{xy}$ should have the fixed ratio 2/T to the diamagnetic moment, using the facts that the velocity of vortex motion is v/c=E/B, and that the interaction energy with the B-field is B·M/2.

To understand the behavior in the vortex liquid phase in the high Tc materials it is essential to realize that it has two separate dynamical aspects, the dynamics of the order parameter and the dynamics of the quasiparticles. Actually, this is also true of ordinary superconductors, but for them, in many situations, the separation between the two is quite simple. For instance, everywhere but close to the transition one neglects thermal fluctuations of the order parameter, to zeroth order, justifying this by the smallness of the region defined by the Ginzburg criterion; then the order parameter is calculated from the temperature-dependent gap equation for quasiparticle energies. In more sophisticated work it is assumed that the fluctuations in amplitude and phase are Gaussian and relatively slow, and are controlled by the pair susceptibility as calculated from the BCS gap equation and inserted into a Ginsburg- Landau effective free energy. The well-known expressions for fluctuation effects result, and in a magnetic field the Abrikosov lattice.

But the x-y model does not have amplitude fluctuations of the magnitude of the spins, only of phase: the spins are not "soft" but are fixed. In fact, in liquid He, which is the "poster child" for x-y critical behavior, Williams[6] has shown that it can be calculated neglecting longitudinal fluctuations entirely and restricting the current J=$\rho_s\nabla\phi$ to be divergenceless, in which case the phase is entirely characterized by a tangle of vortex lines—hence, a "vortex liquid." The longitudinal supercurrents mix with the phonon degrees of freedom which have no critical behavior.

In two-dimensional He films the vortex scenario is well verified[7] as the Kosterlitz-Thouless theory of the phase transition, which assumes from the start that the motions of vortex points are the relevant degrees of freedom. At Tc pairs of vortices unbind and one observes a sharp disappearance of superfluidity, and a dissipation peak. Recent simulations of S Raghu[8] on the 2D x-y model show that the regime above Tc has an additional feature not noticed in previous work: the rotational susceptibility (diamagnetic susceptibility in superconductors) and the Nernst effect are strongly non-linear, which is caused by the fact that the currents due to the *additional* vortices caused by rotational flow are not screened out by the thermally excited vortex pairs. Above Tc, although the vortex pairs are unbound, the long-range (logarithmic) vortex interactions are still so strong that the fluctuations in net vortex density are suppressed, causing the fluid to be *incompressible* as far as net vortex density is concerned. The effect depends on the fact that the amplitude of the boson wave function is everywhere finite except at the vortex centers, and would not appear if the fluctuations were simply like Gaussian fluctuations of soft spins.

This may be understood by considering the effective Hamiltonian of interacting vortex points. These interactions result from the supercurrents $J_s = \rho_s v_s = \rho_s \nabla\varphi$ circulating around the vortex points, which lead to an energy $E_k = \rho_s(\nabla\varphi)^2/2$. Because $\nabla\phi$ falls off only as 1/r with distance, the interactions of the vortex points are proportional to $\rho_s q_i q_j \ln[(R/r_i-r_j)]$, where $q=\pm 1$ is the sign of the quantized vorticity and R is the sample size. The lnR terms drop out if the total vorticity is zero AND if one includes the self-energy $\rho_s \ln(R/a)$, a being a core radius, of each vortex. Except for this cancellation, the self-energy is normally ignored in K-T theory, since only systems with zero total vorticity are considered; otherwise, neglecting it is incorrect.

As was first observed by Onsager, a uniform density of vortices of the same sign mimics uniform rigid rotation except at short distances, and in a rotating sample (or, correspondingly, in a superconducting sample in a magnetic field) the large logarithms may be cut off at the appropriate lattice spacing to match the rotational velocity (or B-field in the superconducting case). Beyond the intervortex distance the extra energy is provided by whatever is causing the rigid rotation or B-field. But the local rotation of the fluid around the vortex cores is not part of the uniform rotation, and costs extra energy—this in fact is the reason for the diamagnetic moment of an Abrikosov superconductor proportional to $\ln(B_c/B)$. It has been thought that above Tc the current of the extra vortices was screened out by the sea of thermal vortices, but this is incorrect. The simplest way to see this is to realize that the extra vortices are not privileged in any way and are just part of the sea of unbound vortices, none of which are bound to any particular other.

None of the above reasoning depends essentially on the two-dimensionality of the model. As Williams showed in reference [6], the 3D phase transition is equally a vortex proliferation transition. The argument for incompressibility depends only on the logarithmic self-energy of the field vortices and on the crucial fact that the B-field adds extra vortices which nonetheless cannot be distinguished in principle from the thermally excited loops.

In the vortex liquid phase in the cuprates the situation is more complicated. The physics is not simply the transition from BCS to Bose condensation of preformed pairs, such as can be nicely modeled with cold atoms and a Feshbach resonance[9], where the only parameter is the ratio of pair binding energy to the Fermi energy. First, the pair binding energy—the gap—must be compared not to the mean field band energy ~t but to the renormalized kinetic energy $g_t t$, where $g_t$ is proportional to doping x. Second, the "spin-charge locking"[10] mechanism is the main reason for the extent of the vortex liquid region. The locking of the charge triad to the spin triad describing the RVB ordering is equivalent to pair-condensing the holes—"holons"—and implies a boson pair field whose current is given by the gradient of phase. The hole pairs are effectively bound by an energy even greater than the observed superconducting gap. A third problem is that the gap vanishes at nodal points so there is no clear separation in frequency space between fluctuations in the order parameter and quasiparticle energies.

These problems also open up an opportunity: to treat the dynamics separately for the order parameter and for the quasiparticles, as is done with ordinary superconductors, but to take into account the interaction of the two kinds of fluctuations. A crude attempt at a theory of this kind was given in my previous treatment[11]. There, I made the approximation that the currents would be predominantly supercurrents, since the relaxation rate for quasiparticles is extremely fast; but this is not at all crucial since at least for transverse currents the normal state diamagnetism and Nernst effect due to quasiparticles are measurable, and they are smaller and strictly linear in B. What I did was to assume that as either T or B was raised, that those portions of the Fermi surface with gap $\Delta$ less than thermal energy kT, or with $H_{c2}(\Delta)<B$, would no longer contribute to the supercurrent. This gives the susceptibility and Nernst effect with the characteristic nonlinear behavior observed by Ong et al.[12] In reference [11] I assumed that the vortex fluid is in the high-temperature regime where kT>E_c, E_c being the vortex core energy, so that the density of vortices is limited only by the available thermal energy; but this is inessential to the calculation there because one can always insert the measured resistivity and then the coefficient is independent of the model for the dissipative behavior.

Another related phenomenon that is observed in the vortex liquid regime is the so-called "Fermi Arc" phenomenon, recently described in detail by Campuzano[13] and previously remarked by him and others. The observation, from ARPES studies of the evolution of the energy distribution functions with rising temperature, is that above Tc the gap disappears progressively starting at the nodes, so that at T>Tc, in k-space directions near the nodes, there appear to be pure Fermi distributions $f(\varepsilon_k/T)$. If the gap is mapped in k-space, it remains essentially unchanged at its antinodes but falls to zero along an arc near the nodes. Quantitatively, Campuzano estimates that for $\Delta(k)>T$, the gap remains unchanged, but $0<\Delta<T$ does not occur above Tc. (One should note that the accuracy does not allow measurements in the critical region around Tc.)

Above Tc our observations above show that pairing still exists, but that the phase of the gap has become time and space dependent. The simple theory of reference [4] argues in fact that the correlation time for the current, or equivalently for the phase of the gap, is given by $\langle \phi(t)\phi(0) \rangle \approx e^{-t/\tau}$, where $\hbar/\tau \approx kT$. This suggests immediately that the reason for the ineffectiveness of the smaller energy gaps is the time-honored phenomenon of "motional narrowing"[14]. The effective Hamiltonian for a quasiparticle of momentum k in the presence of a time-variable energy gap is

$$H = \begin{pmatrix} \omega - \varepsilon_k & \Delta_k(t) \\ \Delta^*_k(t) & \omega + \varepsilon_k \end{pmatrix} \quad [1]$$

The time-development matrix which results from this time-dependent Hamiltonian is

$$U(t) = T\{e^{i\int H(t)dt}\} \approx e^{i\int H(t)dt} \qquad [2]$$

T{} is the time-ordering operator, and the second equality of [2] results because the commutator of H(t) with H(t') is small both for large and for small ε. If Δ is large compared to its relaxation rate, the Fourier components of [2] will cluster about the quasiparticle energy

$$E_k = \sqrt{\varepsilon_k^2 + \Delta^2}$$ ; if Δ is small, the gap will average out before it can have much effect on U(t). This effect has been worked out in detail in the spectral broadening literature[15] and I need only refer to that. Thus if, indeed, the time-scale of vortex motion is determined by kT, as is also suggested by the infrared conductivity in the vortex liquid state, Campuzano's effect would seem to be explained.

A second observation common to Campuzano's work as well as to tunneling studies by Yazdani[16] is that at higher temperatures, even after a considerable fraction of the Fermi surface has become gapless, the maximum gap at the antinodes has hardly decreased. This too is understandable in terms of the "locking" theory: the gap opens up in the vicinity of T* as a self-consistent structure (an RVB) in the spin subspace, which initially has no explicit influence on the "holon" motions, but its magnitude becomes essentially fixed above To. Near To the holons "confine", pairing up and becoming charged quasiparticles. The motivation for their pairing is the preexisting RVB gaps, both Δ and ζ. The anomalous self-energy Δ can be manifest as a true superconducting gap only if it is large enough to overcome the thermal fluctuations in the charge subspace; but it does not fluctuate much in the spin subspace.

In conclusion, we are able to create a consistent account of the dynamics of the vortex liquid which occupies the lower-temperature part of the hitherto mysterious pseudogap region of the cuprate phase diagram. This vortex liquid has no broken symmetry which can be described in terms of a local order parameter, but does have a kind of "topological" ordering expressed by the constraint that the relevant dynamics is that of a phase field. More speculatively, we also propose that the higher-T region is an RVB spin liquid, decoupled from any kind of U(1) charge dynamics.

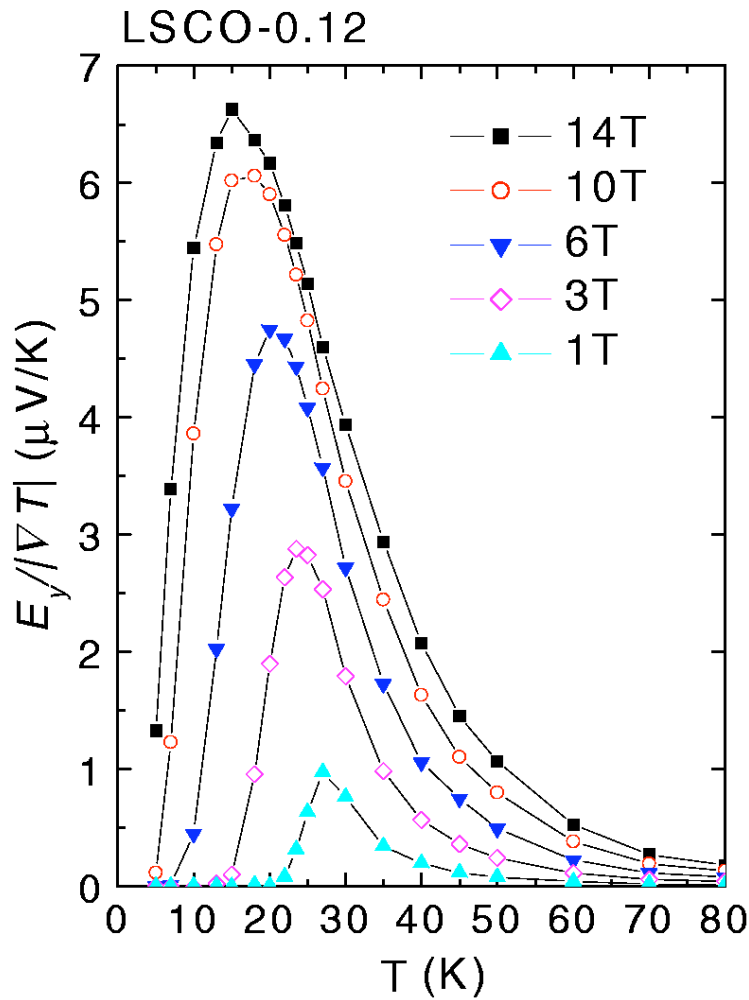

**Figure 1 Nernst signal as a function of T with H field as parameter: underdoped LSCO**

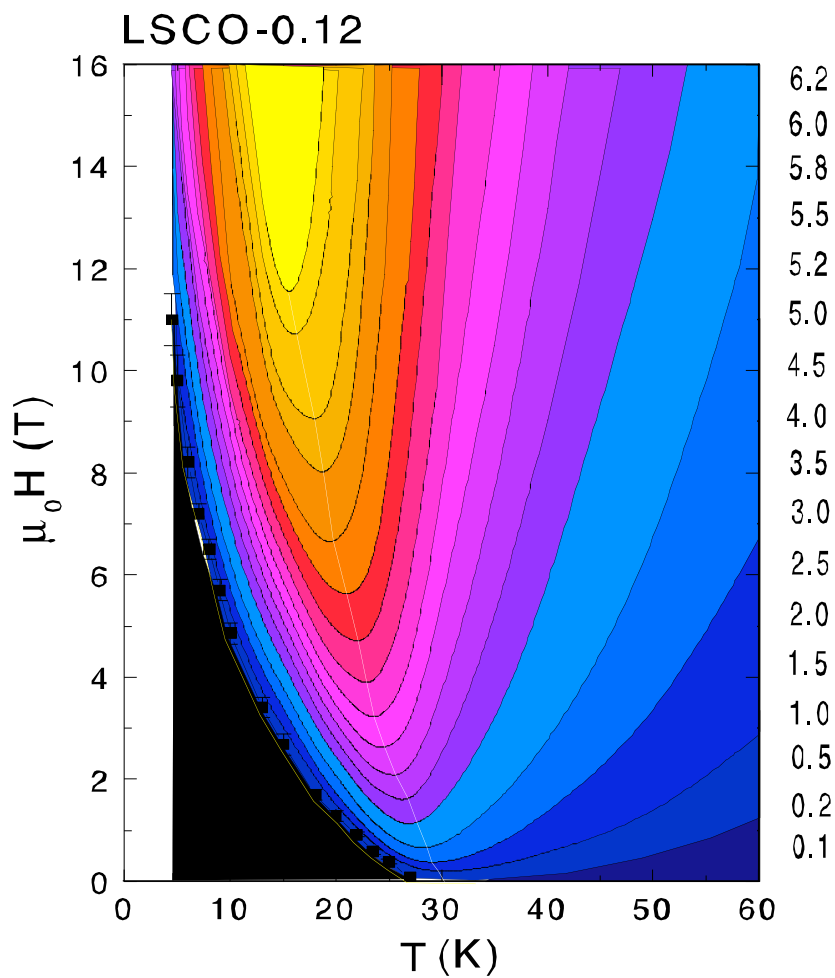

**Figure 2 contours of Nernst signal in H-T Plane**

---

[1] M B Salamon, J Shi, N Overend and M A Howson, Phys Rev B47, 5520 (1993)
[2] J Corson, J Orenstein, et al, Nature 398, 221 (1999)
[3] Yayu Wang et al., Phys. Rev. B **64**, 224519 (2001); Yayu Wang et al., Phys. Rev. Lett. **95**, 247002 (2005); Yayu Wang, Lu Li and N. P. Ong; Phys. Rev. B **73**, 024510 (2006).
[4] P W Anderson, "Two New Vortex Fluids", cond-mat/0606429; tbp, Nature Physics
[5] Y Wang, PhD thesis, Princeton, 2004
[6] G A Williams, Phys Rev Lett 96, 1927 (1987)

[7] D J Bishop and J D Reppy, Phys Rev B22, 5171 (1980)
[8] S Raghu, thesis Princeton 2006
[9] G B Partridge, K E Strecker, R I Kamar, M W Jack, and R G Hulet, Phys Rev Lett 95, 020404 (2005)
[10] P W Anderson, Phys Rev Lett 96, 017001 (2006)
[11] PWA, "Dynamics of the Vortex Fluid in Cuprate Superconductors: the Nernst Effect", cond-mat/0603726
[12] Lu Li et al, Europhys Lett 72, 456-7 (2005)
[13] J-C Campuzano, Nature Physics 2, 447 (2006)
[14] N Bloembergen, E M Purcell, and R V Pound, Phys Rev 73, 679-712, (1947)
[15] P W Anderson, J Phys Soc Japan 9, 316-339 (1954)
[16] A Yazdani, private communication